\documentclass{aa}  
\usepackage{graphicx}
\usepackage{txfonts}
\usepackage{lipsum}
\usepackage{subcaption}        
\usepackage[version=4]{mhchem}
\usepackage{lscape}
\usepackage{placeins}
\usepackage{natbib}
\bibpunct{(}{)}{;}{a}{}{,}
\usepackage{booktabs}
\usepackage{threeparttable}
\usepackage{siunitx}
\usepackage{stfloats}
\usepackage{textcomp} 
\usepackage{makecell}
\usepackage{bm}
\usepackage[colorlinks=true, linkcolor=blue, citecolor=blue, urlcolor=blue]{hyperref}

\begin{document}

\titlerunning{A dataset of 144,111 PAH IR spectra}
\authorrunning{Mai et al.}

\title{A machine-learning-driven dataset of 144,111 polycyclic aromatic hydrocarbon infrared         spectra for interpreting aromatic infrared bands}

   \author{Xinghong Mai \inst{1}
   \and Zu-Jia Lu \inst{1} 
   \and Zhao Wang \inst{1,2}\fnmsep\thanks{Corresponding author: \email{zw@gxu.edu.cn}}
   }

   \institute{Laboratory for Relativistic Astrophysics, Department of Physics, Guangxi University, 530004 Nanning, China
   \and Center for Applied Mathematics of Guangxi (Guangxi University), 530004 Nanning, China
   }

\date{Received \today}

\abstract
{The interpretation of interstellar aromatic infrared bands (AIBs) relies on reference spectral libraries, but the chemical space covered by existing databases remains incomplete relative to the vast structural diversity of polycyclic aromatic hydrocarbons (PAHs) expected in the interstellar medium.}
{We present a large-scale, high-throughput PAH infrared (IR) spectral database that complements existing reference libraries by providing significantly more continuous coverage of the PAH chemical space for small to medium-sized PAHs.}
{We used a machine-learning double-harmonic (ML-DH) framework that integrates neural network force fields with electronic dipole models. This framework achieves quantum-chemical accuracy at a fraction of the computational cost. It has been validated against independent density functional theory calculations and benchmark spectral databases.}
{The resulting dataset comprises $144,111$ IR spectra for $48,037$ closed-shell, even-carbon benzenoid PAH structures across neutral, cationic, and anionic charge states, complete with per-spectrum uncertainty estimates.}
{The spectral library, optimized geometries, and model code are publicly available. They can be used to complement existing libraries for interpreting and modeling JWST AIB observations and those from future IR missions.}


\maketitle
\nolinenumbers

\section{Introduction}

Polycyclic aromatic hydrocarbons (PAHs) are widely recognized as key molecular species that are abundant throughout the interstellar medium (ISM; \citealt{leger1984, allamandola1985}). Their infrared (IR) emission features, known as aromatic infrared bands (AIBs), are powerful probes of local physical and chemical conditions, as their spectral signatures are highly sensitive to factors such as the ionization state, radiation field, and molecular size \citep{peeters2002, galliano2008, maragkoudakis2020, rigopoulou2021, li2024}. However, the effective use of PAHs as diagnostic tools is severely hampered by their vast structural diversity, which gives rise to an exceptionally complex structure-spectral relationship \citep{peeters2021,li2020}. For example, the number of possible benzenoid isomers alone exceeds one million for molecules with 60 or fewer carbon atoms, with more than 300,000 cata-condensed isomers existing for \ce{C54H30} \citep{Dias2004, Dias2007}. Interpreting the observed AIBs thus demands sufficiently large and representative reference spectral libraries that can adequately sample this immense chemical space.

The current benchmark resource, the NASA Ames PAH IR Spectroscopic Database (PAHdb), provides 10,749 theoretical and 84 experimental spectra collected over more than 15 years \citep{bauschlicher2010, boersma2014, bauschlicher2018, mattioda2020, ricca2026}. Despite its invaluable role, this library covers only a fraction of the expected interstellar PAH diversity. The prohibitive computational cost of traditional quantum chemical calculations (QCCs) remains the primary bottleneck for generating reference spectra at the scale required to match the known structural complexity. To complement this benchmark by providing denser coverage, we present a large-scale spectral dataset containing 144,111 IR spectra for 48,037 small to medium-sized PAH structures in neutral, anionic, and cationic charge states.

A key challenge in constructing such a dataset is balancing high throughput with quantum-chemical accuracy. End-to-end machine learning (ML) models offer substantial speedups \citep{kovacs2020, meng2021, mcgill2021, saquer2024, he2026, liu2026}, yet their predictions often degrade for molecules outside their training distribution. Conversely, ML-based molecular dynamics methods incorporate physical principles by learning potential energy surfaces \citep[PESs;][]{gastegger2017}, but they suffer from systematic errors in high-frequency modes and require expensive trajectory simulations \citep{tang2023, bhatia2025}. Here, we adopt an ML-based double-harmonic (ML-DH) model to bypass these limitations. By directly predicting vibrational frequencies and IR intensities from equilibrium structures via charge-specific neural networks trained on first-principles forces and dipoles, the ML-DH model achieves near-QCC accuracy at a fraction of the cost, enabling high-throughput spectral generation across vast PAH chemical space.

Using the ML-DH framework, we generated high-throughput IR spectra for 48,037 unique PAH structures drawn from the Computational Database of Polycyclic Aromatic Systems (COMPAS) database \citep{wahab2022, mayo2024, wahab2024} in three charge states. Rigorously validated against independent QCCs and existing spectral libraries, both the dataset and model code are made publicly available to facilitate the interpretation of AIBs observed by the \textit{James Webb} Space Telescope (JWST) and advance our understanding of the molecular composition and evolution of the ISM.

\section{Methodology}

\subsection{Overview and core approximation}
\label {DH}

The theoretical foundation of the ML-DH model rests on the double harmonic (DH) approximation \citep{neugebauer2002}, wherein the vibrational spectrum of a molecule is uniquely determined from its equilibrium geometry by two fundamental properties:
\begin{enumerate}
    \item Harmonic vibrational frequencies ($\nu$) and normal modes ($\bm{Q}$), derived from the second derivatives of the PES ($E$) with respect to atomic displacements;
    \item IR intensities ($I$), governed by the first derivatives of the molecular dipole moment ($\bm{\mu}$) with respect to the normal coordinates ($\bm{Q}$).
\end{enumerate}

Under this framework, the Taylor expansion of the potential energy ($E$) about the equilibrium geometry is truncated at second order. The derivation begins with the dynamical matrix ($\bm{D}$), defined as the mass-weighted Hessian of $E$. For an $N$-atom system with $3N$ Cartesian degrees of freedom indexed by $i, j = 1, \dots, 3N$, the matrix elements ($D_{ij}$) are formulated as

\begin{equation}
\label{Eq1}
\bm{D}_{ij} = \frac{1}{\sqrt{m_i m_j}} \frac{\partial^2 E}{\partial r_i \partial r_j},
\end{equation}

\noindent  where $r_i$ and $r_j$ are the corresponding Cartesian coordinates, and $m_i$ and $m_j$ are the masses of the atoms to which these degrees of freedom belong.

$\bm{D}$ is then diagonalized to obtain the frequencies ($\nu$). Solving the eigenvalue problem,

\begin{equation}
\label{Eq2}
\bm{D} \bm{Q}_p = \lambda_p \bm{Q}_p,
\end{equation}

\noindent where $\lambda_p$ is the eigenvalue and the normal mode ($\bm{Q}_{p}$) describes the atomic displacement pattern of the $p$-th vibration. The frequency ($\nu$) of mode $p$ is given by $\sqrt{\lambda_{p}}/2\pi$.

The intensity of the $p$-th vibrational mode is proportional to the square of the derivative of $\bm{\mu}$ with respect to $Q_{p}$:

\begin{equation}
\label{Eq4}
I_{\nu_p} = \frac{1}{4\pi\varepsilon_0} \frac{N_{\mathrm{A}}\pi}{3c^2} \left| \frac{\partial \bm{\mu}}{\partial Q_p} \right|^2,
\end{equation}

\noindent where $N_{\mathrm{A}}$ is Avogadro's constant, $c$ is the speed of light, and $\varepsilon_0$ is the vacuum permittivity. The derivative ($\partial \bm{\mu} / \partial Q_p$) is evaluated via the chain rule from Cartesian dipole derivatives:

\begin{equation}
\label{Eq5}
\frac{\partial \mu_\alpha}{\partial Q_p} = \sum_{j=1}^{3N} \frac{\partial \mu_\alpha}{\partial r_j} \frac{\partial r_j}{\partial Q_p},
\end{equation}

\noindent where $\alpha \in \{x,y,z\}$ denotes the Cartesian component of $\bm{\mu}$.

The DH approximation rests on two core assumptions: truncating the PES expansion at second order, and retaining only the linear term in the Taylor expansion of the molecular dipole moment. For PAHs, these assumptions are generally well justified across mid- and far-IR vibrational modes, where anharmonic effects remain modest; however, C-H stretching modes can exhibit non-negligible anharmonicity \citep{langhoff1996}. Despite this limitation, the DH framework remains the established standard for large-scale PAH spectroscopic databases, such as PAHdb, primarily owing to its computational efficiency, which enables high-throughput spectral generation across the vast PAH chemical space.

\subsection{Machine learning implementation}

Conventional QCCs rely on computationally expensive electronic structure calculations to evaluate the PES and dipole moment ($\bm{\mu}$) surface, severely limiting their scalability to large molecular datasets. To circumvent this computational bottleneck, we replaced these calculations with two ML models.

The first model is a neural network force field (NNFF) designed to predict the PES \citep{montes-campos2021, pan2024}. The NNFF combines spherical Bessel descriptors with a deep residual network architecture to encode local atomic environments, achieving high fidelity in reproducing density functional theory (DFT) energies and forces. These ML-predicted potential energy derivatives enable geometry optimization, Hessian matrix construction via automatic differentiation (Eq.~\ref{Eq1}), and subsequent normal mode analysis (Eq.~\ref{Eq2}).

The second model is an electron-passing neural network (EPNN) trained to predict $\bm{\mu}$ \citep{metcalf2021}. Employing a charge-conserving message-passing graph architecture, the EPNN directly predicts $\bm{\mu}$ without resorting to ambiguous atomic partial charge assignments. This capability allows the evaluation of Cartesian dipole derivatives via finite differences (Eq.~\ref{Eq5}), which are combined with the normal modes derived from the NNFF to yield harmonic IR intensities (Eq.~\ref{Eq4}). Although the independent training of the NNFF and EPNN on common DFT reference data raises potential concerns regarding error propagation, our empirical validation in the following section demonstrates that such interaction errors remain minor in practice.

\begin{figure}
\centering
\includegraphics[width=0.5\textwidth]{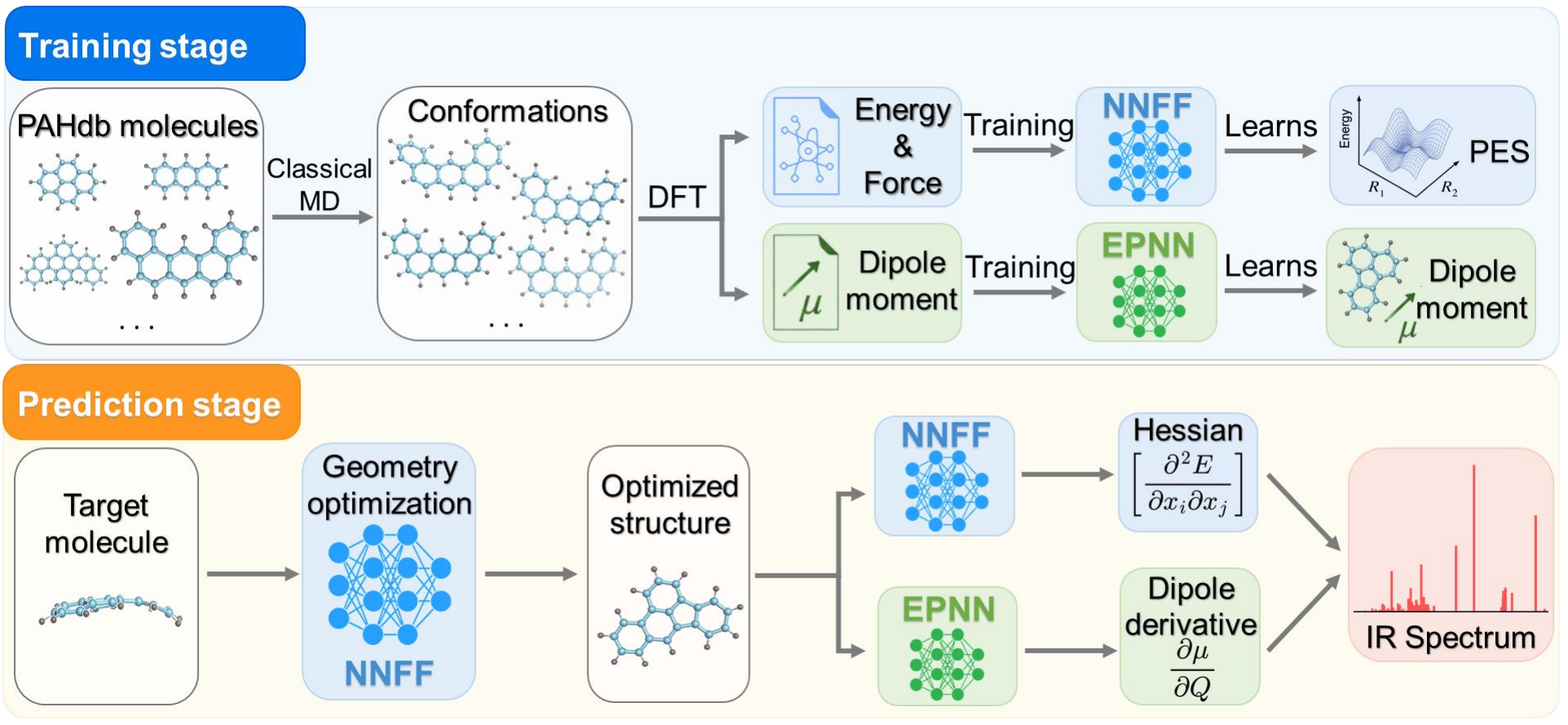}
\caption{Overview of the ML-DH framework. Top: Training stage, using DFT energies, forces, and dipoles to train the NNFF and EPNN. Bottom: Prediction stage, yielding IR spectra via NNFF optimization, normal mode analysis, and EPNN dipole derivatives.}
\label{F1}
\end{figure}

Figure~\ref{F1} illustrates the workflow of the ML-DH framework. In the training stage, single-point DFT calculations generated across a diverse conformation set provide reference energies, atomic forces, and dipole moments to train the NNFF and EPNN in parallel. In the prediction stage, the framework executes three sequential steps: (i) geometry optimization using NNFF-predicted energies and forces to locate equilibrium structures; (ii) normal mode analysis via automatic differentiation of the NNFF to derive harmonic wave numbers; and (iii) evaluation of EPNN-derived dipole derivatives to extract IR intensities. The predicted spectra are subsequently compiled for each molecule across all three charge states.

\subsection{Data construction and model training}

To tailor the NNFF and EPNN models to the DH framework for the PAHs investigated here, we implemented a charge-state-specific training strategy, constructing separate models for neutral, cationic, and anionic species. For the neutral model, we used 651 representative PAH structures from PAHdb (retrieved from our previous study, \citealt{mai2025}), anchoring our ML models to the same chemical foundation as the benchmark. To generate a diverse training ensemble for each charge state, we performed classical molecular dynamics simulations starting from these parent structures, supplemented by random Gaussian spatial perturbations applied to the equilibrium geometries. This protocol produced a total of 16,275 unique molecular conformations. For the charged models, this identical set of 16,275 conformations was retained, with supplementary single-point DFT calculations carried out independently for the $+1$ and $-1$ charge states.

All quantum chemical reference calculations were performed using Gaussian 16 \citep{Frisch2016} at the B3LYP/4-31G level of theory, a setup widely adopted for large-scale PAH IR spectral computations due to its high computational efficiency. Because this level of theory systematically overestimates bond stiffness, leading to higher vibrational wave numbers, we applied the empirical scaling factors from \citet{bauschlicher2018}: 0.960 for the C--H stretching region ($>2500~\text{cm}^{-1}$), 0.952 for the C--C stretching and C--H in-plane bending region ($1111.1\text{--}2500~\text{cm}^{-1}$), and 0.956 for the C--H out-of-plane bending region ($<1111.1~\text{cm}^{-1}$), which were calibrated against experimental gas-phase spectra. 

We acknowledge that the 4-31G basis set lacks diffuse and polarization functions, which can lead to an underestimate of absolute dipole moments, particularly for anionic PAHs. Nevertheless, we adopted this level for its computational efficiency in high-throughput generation. To quantify the impact of basis set incompleteness, we benchmarked a representative subset of 200 training molecules at the B3LYP/6-31G(d) level. The results demonstrate high consistency, yielding average minimal mean absolute errors (MAEs) of $8.28\times10^{-4} \text{ Hartree/Bohr}$ for atomic forces, and $0.038\text{ Debye}$ for dipole moments, with all coefficients of determination exceeding $R^{2}>0.995$. This limitation primarily slightly affects absolute intensities, whereas relative intensity patterns and peak positions remain robust after empirical frequency scaling \citep{bauschlicher1997}. Consequently, we focused on closed-shell, even-carbon benzenoid PAHs, which are the most computationally robust class where multi-reference effects are minimal.

\subsection{Model training}
\label{s-MT}
Using the charge-state-resolved datasets described above, we trained dedicated NNFF and EPNN models for each of the three charge states. The NNFF model employs a four-layer ResNet as its neural network backbone, with a core-width sequence of 64--32--16--16 and a training batch size of 4096. The EPNN model utilizes a charge-conserving message-passing graph neural network architecture, incorporating three message-passing layers with hidden dimensions of (32, 32, 16) and a batch size of 2560. Both models were trained using the VeLO optimizer. To reduce sensitivity to outlier errors during optimization, the training loss was formulated using the log-cosh function. For all charge states, the datasets were partitioned into training and validation sets using a 0.8/0.2 split.

Prior to vibrational analysis, PAH molecular geometries were optimized using the Broyden-Fletcher-Goldfarb-Shanno algorithm as implemented in the Atomic Simulation Environment \citep{larsen2017}, with potential energies and atomic forces supplied directly by the trained NNFF. Optimization convergence was defined by a maximum atomic force threshold of $0.0005~\mathrm{eV/\text{\AA}}$, subject to a maximum limit of 1000 steps. Upon achieving the relaxed geometry, the Hessian matrix was constructed via automatic differentiation of the NNFF PES. To isolate intrinsic vibrational motions, the Hessian matrix was mass-weighted, and translational and rotational degrees of freedom were projected out, restricting matrix diagonalization strictly to the $3N-6$-dimensional vibrational subspace. During the normalization of projection vectors, a numerical threshold was applied to vectors with norms below $10^{-15}$ to prevent instability from division by zero. Diagonalization of the projected, mass-weighted Hessian yielded the harmonic vibrational frequencies and corresponding normal modes.

\begin{figure}
\centering
\includegraphics[width=0.5\textwidth]{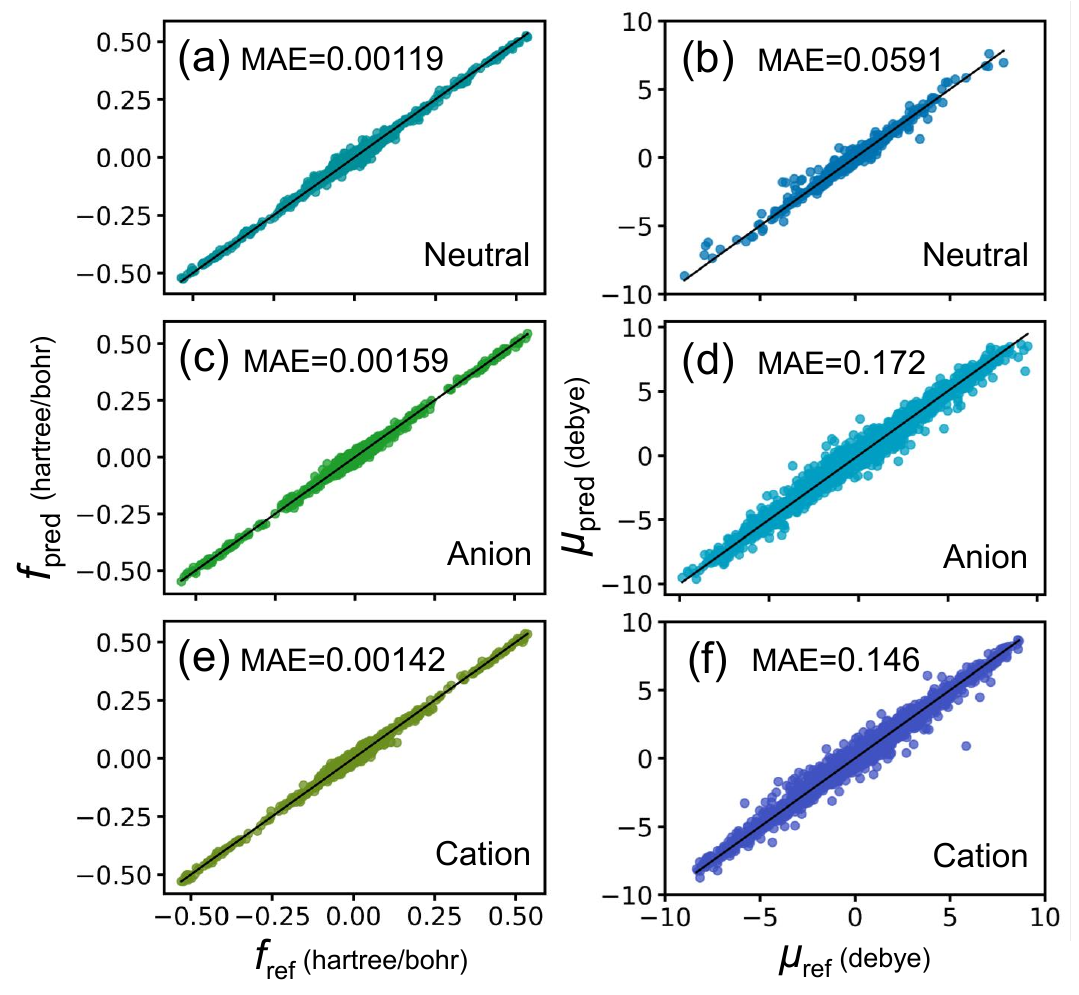}
\caption{Comparison of ML-predicted and DFT-calculated reference values for three charge states.  Panels (a), (c), and (e): NNFF-predicted atomic force components, $f$.  Panels (b), (d), and (f): EPNN-predicted molecular dipole moment components, $\mu$.  The MAE is reported in Hartree/Bohr for $f$ and Debye for $\mu$.}
\label{F2}
\end{figure}

To rigorously evaluate the performance of our core ML components, an independent test set comprising 20\% of the 16,275 conformations was held out from the training process. Figure~\ref{F2} compares the atomic forces predicted by the NNFF and dipole moments predicted by the EPNN against their benchmark DFT values across all charge states for these unseen conformations. Both surrogate models demonstrate high predictive accuracy across neutral and ionized species, closely reproducing their respective targets with MAEs. This strong agreement confirms the capacity of the learned representations to capture the distinct PESs and electronic polarizabilities inherent to different ionization states. Consequently, these validated neural network models provide the high-fidelity PES and dipole predictions required to drive the full ML-DH spectral synthesis described in Sect.~\ref{DH}.

For large-scale spectral generation, the computational cost per molecule is dominated by geometry optimization (typically requiring $\sim 400$ iterations). Training the NNFF and EPNN models required approximately 57 hours on a NVIDIA RTX4090 GPU. Once trained, generating the full database of 144,111 spectra across the three charge states consumed approximately 45,492 CPU-hours. This performance corresponds to a speedup of approximately 36 times compared to conventional DFT calculations, which would require an estimated 1,637,028 CPU-hours on the same machine. This baseline DFT cost was estimated by weighting size-dependent average DFT runtimes from a COMPAS test dataset across the size distribution of our database.

A legitimate concern is that the training set comprises only 651 parent structures, whereas the final dataset includes 48,037 COMPAS structures. 
However, vibrational spectra are intrinsically local: mode frequencies and intensities are determined by local chemical environments, not global topology. For PAHs, the relevant environments, for example edges, bay and fjord motifs, and fused-ring junctions, are finite and well sampled in the training set, especially via conformational perturbations. Validation against the independent COMPAS structures confirms that the model captures this local mapping, enabling reliable generalization despite the modest training set.

\section{Results and discussion}

\subsection{Model validation}

Given that the primary goal of this work is to provide a comprehensive spectral library of PAHs, it is essential to critically assess the accuracy of our ML-DH approach. To this end, we systematically compared the predicted IR spectra against two distinct sets of reference data: (i) 5,166 DFT (B3LYP/4-31G) harmonic spectra for 1,722 PAHs randomly selected from the COMPAS datasets in three charge states and (ii) 4,579 theoretical and 59 experimental spectra from PAHdb v4.0 for hydrocarbon PAHs across different charge states, excluding structures present in the training set. The validation set used here for spectral comparison is independent of the held-out test set in Sect.~\ref{s-MT}, which was used only to validate the predicted forces and dipoles.

\begin{figure}
\centering
\includegraphics[width=0.5\textwidth]{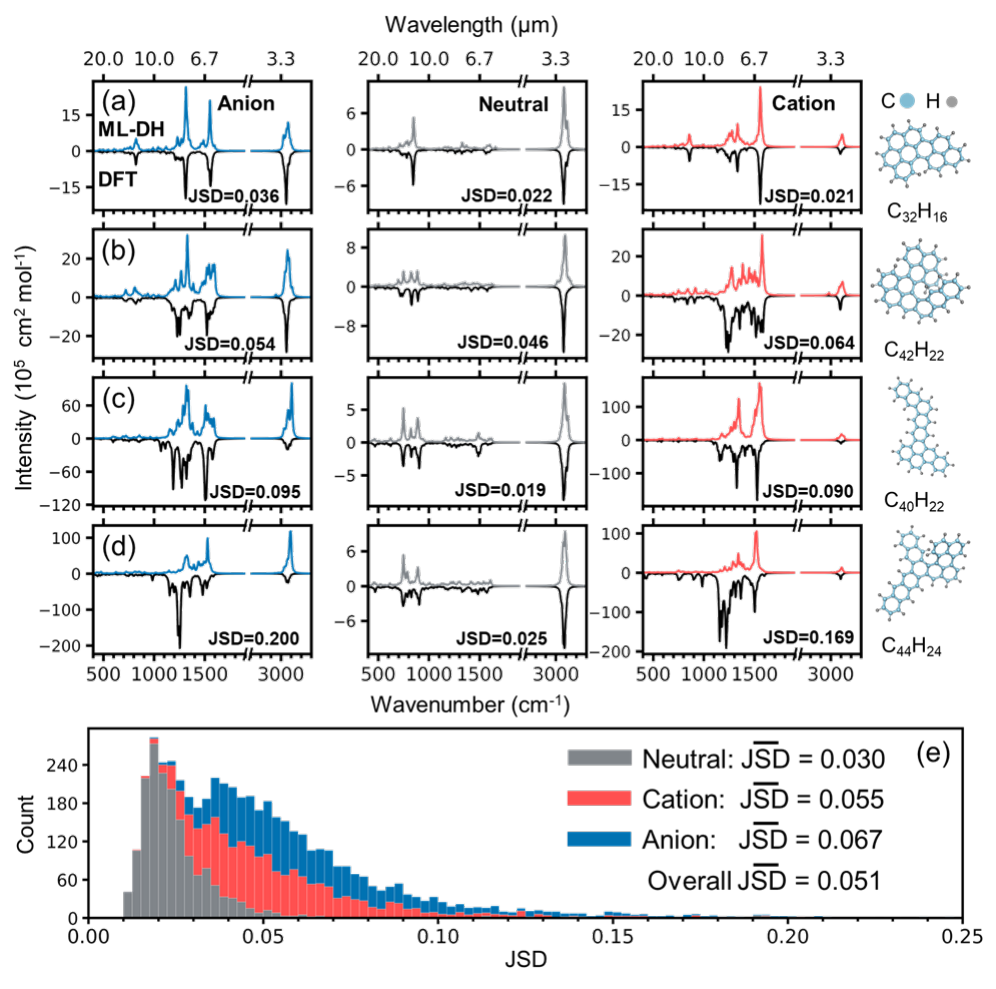}
\caption{Representative examples of ML-DH-predicted spectra compared against DFT reference spectra for selected PAHs. Upper curves denote ML-DH predictions, and bottom mirrored curves denote DFT references. Each row corresponds to a representative molecule, arranged from best (top) to worst (bottom) JSD agreement. The three columns from left to right correspond to anionic, neutral, and cationic charge states, respectively. Panel (e): JSD distribution between ML-DH predicted and DFT reference spectra for 5,166 spectra of 1,722 PAHs across three charge states.
}
\label{F3} 
\end{figure}

We first assessed the ML-DH performance by comparing its predictions against the 5,166 DFT-computed harmonic spectra of COMPAS PAHs excluded from the training set. This validation aims not to test extrapolation to new chemical spaces, but to confirm DFT-comparable accuracy for molecules representative of the COMPAS dataset. Figure~\ref{F3} displays representative comparisons for four PAHs selected to span the full range of agreement quality, from good (row a) to poor (row d). The three columns correspond to anionic, neutral, and cationic charge states, respectively. To quantify spectral similarity, we adopted the Jensen–Shannon divergence (JSD), which is sensitive to both peak positions and shapes, as an advantage over point-wise metrics such as the MAE \citep{tang2026}.

The examples in Fig.~\ref{F3}(a-d) illustrate the relationship between JSD and the visual quality of the spectral agreement. For JSD values below approximately 0.03, such as those observed for neutral \ce{C32H16} (0.022), cationic \ce{C32H16} (0.021), and neutral \ce{C40H22} (0.019), the ML-DH spectra are virtually indistinguishable from the DFT references, faithfully reproducing all major peak positions and relative intensities. In the moderate JSD range of 0.03–0.06, exemplified by anionic \ce{C32H16} (0.036), neutral \ce{C42H22} (0.046), and anionic \ce{C42H22} (0.054), the overall spectral shapes remain well captured, although minor discrepancies in peak intensities or very weak features begin to appear. Even at JSD values around 0.1–0.2, as seen for anionic \ce{C40H22} (0.095), cationic \ce{C44H24} (0.169), and anionic \ce{C44H24} (0.200), the predicted spectra still correctly reproduce the dominant peak positions and the overall band envelope.

Figure~\ref{F3}e presents the JSD distribution for the 5,166 DFT harmonic spectra. The vast majority of spectra show low JSD values: approximately 99.23\% fall below 0.2, and 92.88\% are below 0.1. The mean JSD across the entire dataset is 0.051, confirming good overall agreement between the ML-DH predictions and the DFT reference spectra. Only a small fraction of spectra (0.77\%, barely visible in the figure) yield JSD values exceeding 0.2, a regime where predictions are less reliable.

To further validate the ML-DH model, we compared its predictions against the PAHdb reference, which includes 4,579 theoretical harmonic spectra and 59 experimental spectra for hydrocarbon PAHs in different charge states. The PAHdb theoretical spectra, like ours, are DFT-based harmonic calculations, albeit at different levels of theory (e.g., B3LYP/6-31G(d)). This is therefore a computational-to-computational validation rather than a test against experiment, but PAHdb remains the current benchmark for interpreting AIBs, and we aim to show that our predictions are consistent with it within the typical spread of DFT methods, while offering far broader coverage.

\begin{figure}
\centering
\includegraphics[width=0.5\textwidth]{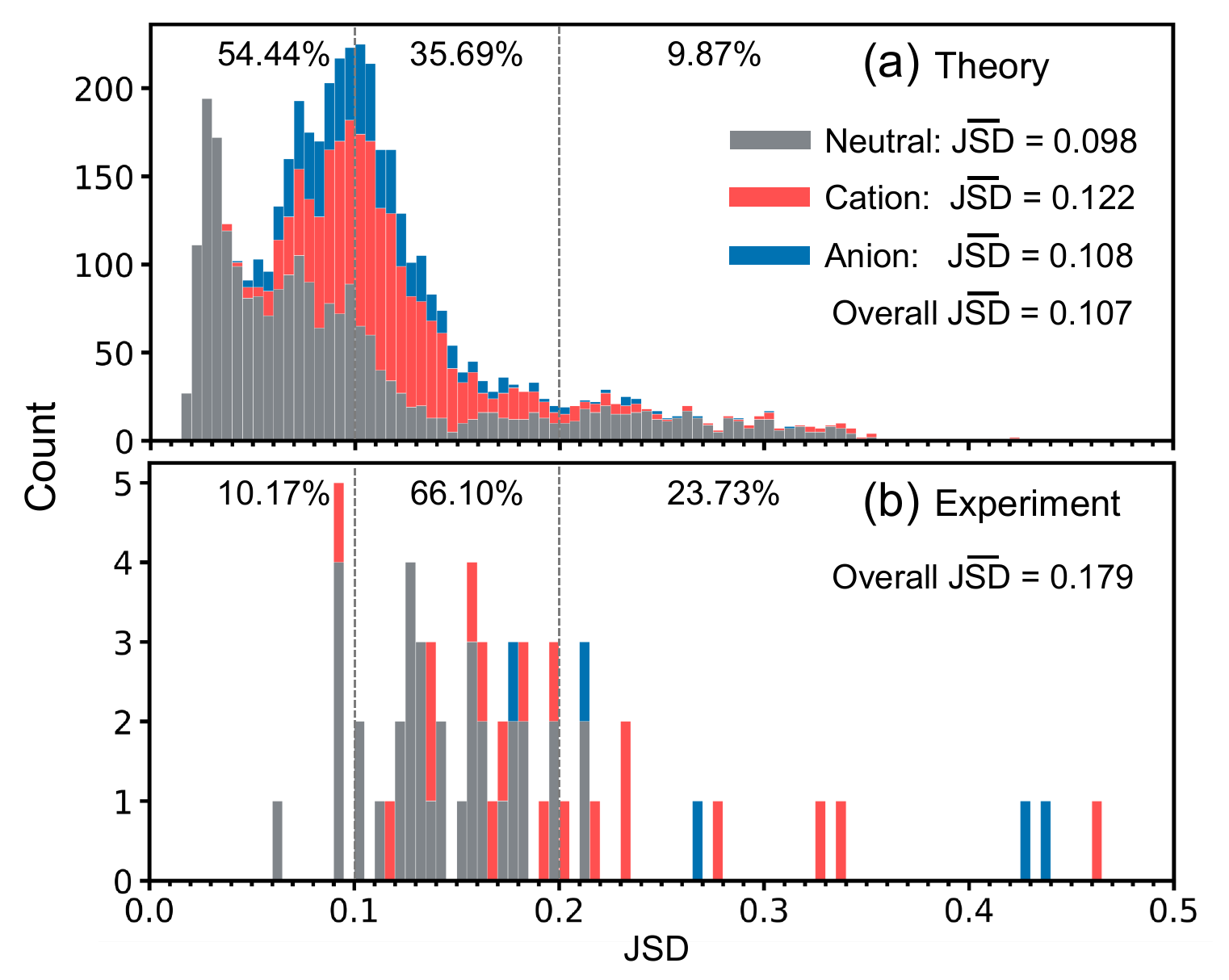}
\caption{JSD distributions comparing ML-DH-predicted spectra against PAHdb (v4.0) reference spectra. Panel (a): Comparison with 4,579 theoretical harmonic spectra (2,392 neutral, 1,507 cationic, and 680 anionic PAHs). Panel (b): Comparison with 59 experimental spectra (35 neutral, 19 cationic, and 5 anionic PAHs). The three charge states are distinguished by different bar shades.}
\label{F4} 
\end{figure}

Figure~\ref{F4} presents the JSD distributions for these comparisons. 
For the theoretical references (panel a), the distribution is heavily concentrated at low values: 90.13\% of all JSD values fall below 0.2, confirming broad compatibility with PAHdb. For neutral PAHs, the distribution peaks sharply at JSD $\sim 0.025$ and is strongly skewed toward low JSD: 86.12\% fall below 0.2, 68.31\% below 0.1, with a mean of 0.098. Cationic PAHs show a broader distribution peaking at JSD $\sim 0.095$, with 93.70\% below 0.2 and 35.10\% below 0.1, yielding a mean of 0.122. Anionic PAHs exhibit the most dispersed distribution: 96.32\% below 0.2, 48.53\% below 0.1, and a mean of 0.108. This charge-state dependence likely arises because ionized PAHs undergo more pronounced charge redistribution, leading to greater sensitivity in dipole moment derivatives and thus posing additional challenges for spectral prediction.

Compared with the internal COMPAS test set (Fig.~\ref{F3}e), validation against PAHdb theoretical spectra yields a higher mean JSD ($0.107$ vs. $0.051$), reflecting differences in structural diversity, size range, and charge-state proportions between the datasets. When restricted to the same size range (number of carbon atoms $N_{\mathrm{C}} < 45$, 1,391 spectra), the mean JSD drops to $0.071$. Overall, $90.13\%$ of the PAHdb spectra exhibit $\mathrm{JSD} < 0.2$ and $54.44\%$ exhibit $\mathrm{JSD} < 0.1$, indicating reliable generalization to an independent benchmark. Although PAHdb calculations predominantly use B3LYP/6-31G(d) while our model was trained on B3LYP/4-31G, the good agreement confirms that empirical frequency scaling aligns peak positions and relative intensities well across these basis sets for neutral species, despite larger discrepancies in charged states.

We next compared the ML-DH predicted spectra with 59 experimental PAH spectra from PAHdb. As shown in Fig.~\ref{F4}b, this comparison yields a higher mean JSD of $0.179$, with neutral, cationic, and anionic PAHs exhibiting mean JSD values of $0.143$, $0.212$, and $0.304$, respectively. Overall, $76.27\%$ of the comparisons score $\mathrm{JSD} < 0.2$ and $10.17\%$ score $\mathrm{JSD} < 0.1$. The larger discrepancies relative to theoretical benchmarks stem primarily from experimental conditions, including finite-temperature anharmonicity \citep{mackie2015, mackie2018, peeters2021} and matrix interactions, which are absent in harmonic $0~\mathrm{K}$ calculations for isolated molecules. Accounting for these effects would require empirical line broadening and frequency shifts that vary with temperature and matrix environment. Nevertheless, the agreement confirms that the ML-DH model reproduces dominant band positions and overall spectral envelopes across different charge states.

\subsection{Large-scale spectral dataset}

Using the validated ML-DH framework, we generated $144,111$ IR spectra for $48,037$ PAH structures from the COMPAS dataset across neutral, cationic, and anionic charge states. The complete dataset is available at \href{https://doi.org/10.5281/zenodo.21506061}{10.5281/zenodo.21506061}, with the model code hosted at \href{https://github.com/zwAstroChem/ML-DH-PAH-IR}{GitHub}.

\begin{figure}
\centering
\includegraphics[width=0.5\textwidth]{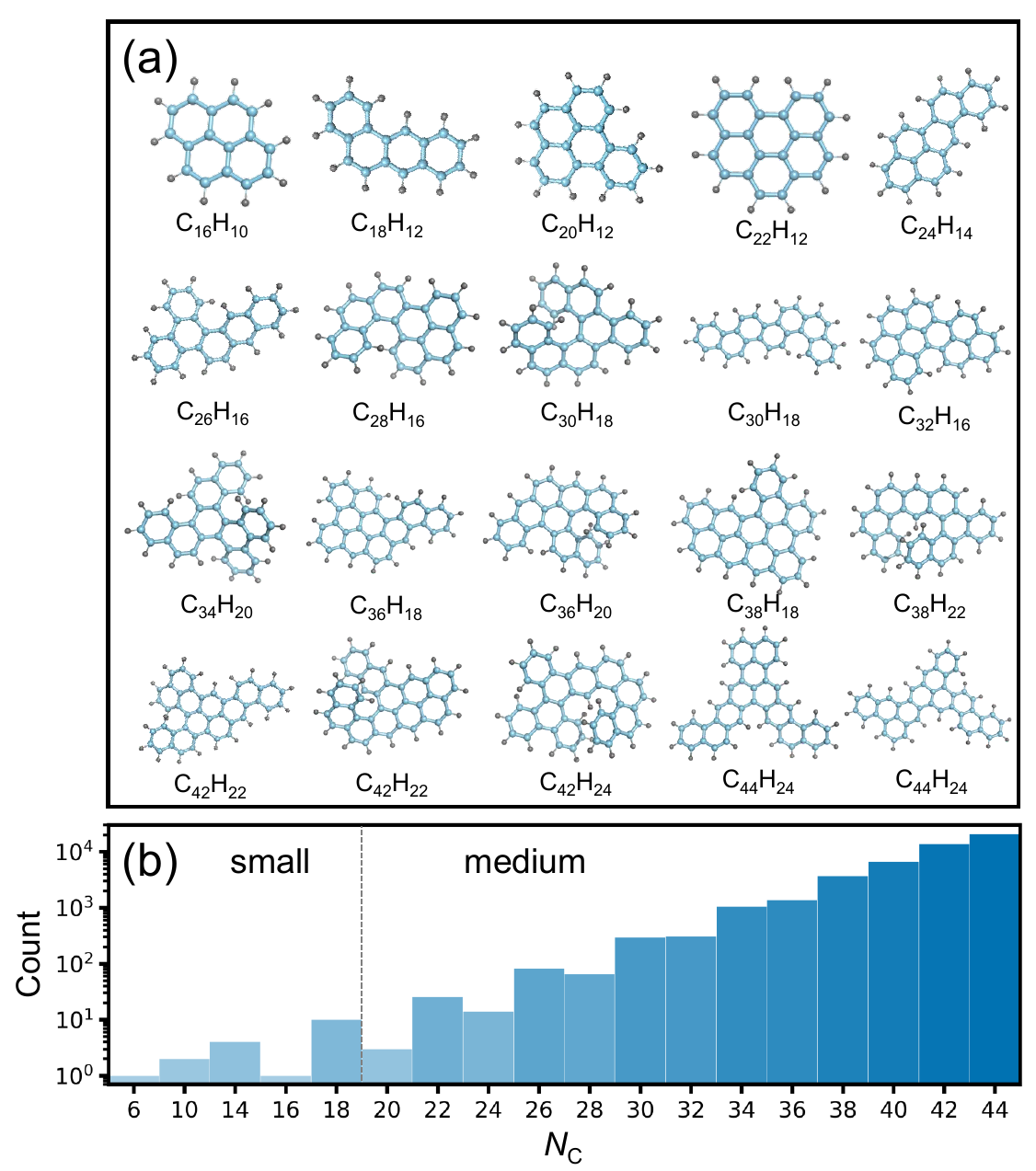}
\caption{Composition distribution and representative structures of the 48,037 PAHs in the dataset. Panel (a): Representative molecular structures across different sizes. Panel (b): $N_{\mathrm{C}}$ distribution of the dataset, spanning 6--44 carbon atoms, with the small ($N_{\mathrm{C}} < 18$) and medium-sized ($18 \leq N_{\mathrm{C}} < 50$) PAH regimes indicated.}
\label{F5} 
\end{figure}

These pure hydrocarbon species span a broad size range and structural variety. As illustrated in Fig.~\ref{F5}a, the dataset incorporates peri-condensed and cata-condensed topologies featuring armchair, zigzag, and bay edges, along with nonplanar geometries distorted by $\pi$-electron stacking or steric crowding. The size distribution ranges from $6$ to $44$ carbon atoms (Fig.~\ref{F5}b), displaying a near sub-exponential increase with $N_{\mathrm{C}}$ ($1$ molecule at $N_{\mathrm{C}} = 6$, $298$ at $N_{\mathrm{C}} = 30$, $6707$ at $N_{\mathrm{C}} = 40$, and $20483$ at $N_{\mathrm{C}} = 44$). This distribution mirrors the combinatorial explosion of PAH isomers with increasing ring count \citep{Dias2004, Dias2007}. Following the conventional classification in the field (e.g., \citealt{boersma2014,ricca2012}), we categorized our PAH structures into small ($N_{\mathrm{C}} < 18$), medium ($18 \le N_{\mathrm{C}} < 50$), and large ($N_{\mathrm{C}} \ge 50$) classes. Our dataset primarily covers the small-to-medium regime, with the majority of structures concentrated in the medium-sized category ($35 < N_{\mathrm{C}} < 45$). We note that the large PAHs ($N_{\mathrm{C}} \ge 50$) are beyond the current scope of our dataset.

The dataset draws from multiple COMPAS installments: COMPAS-3 \citep[$80.7\%$, peri-condensed polybenzenoid hydrocarbons, where a carbon can be shared by up to three rings;][]{wahab2024}, COMPAS-1 \citep[$17.7\%$, cata-condensed polybenzenoid hydrocarbons;][]{wahab2022}, and COMPAS-2 \citep[$1.6\%$, cata-condensed hetero-polycyclic aromatic systems;][]{mayo2024}. The restriction to even-carbon numbers stems from the exclusion of open-shell species across COMPAS sources: COMPAS-3 discarded non-Kekul\'{e}an and radicaloid structures via fractional occupation number weighted density filtering, while COMPAS-1 excluded acenes with more than six linearly annulated rings. Furthermore, we removed molecules containing four-membered rings due to performance degradation in our NNFF and EPNN models, which were trained exclusively on six-membered rings. These screening criteria selectively isolate closed-shell, even-carbon benzenoid PAHs, which are the most computationally robust PAH class. The high proportion of nonplanar species in the dataset is primarily attributed to peri-condensed structures from COMPAS-3, where fjord, cove, or helical motifs induce out-of-plane distortions via steric crowding.

To provide a per-spectrum reliability indicator, we adopted a nearest-neighbor transfer strategy, grounded in the assumption that molecules with similar structures exhibit similar prediction errors due to the smoothness of the ML model's performance across chemical space. For every molecule, we computed its extended‑connectivity fingerprint (radius = 3;  \citealt{rogers2010, meng2023machine}) and evaluated the cosine similarity against all molecules in the independent validation set (5,166 DFT spectra). The JSD of the most similar validation molecule, defined as the one with the highest cosine similarity, is then assigned as a proxy for the expected prediction error of the target spectrum. This approach leverages the validation‑set errors as empirical anchors; a low assigned JSD indicates that the target molecule has a close structural analog with verified ML‑DH accuracy, while a high JSD flags molecules in structurally sparse regions where predictions should be interpreted with caution. Across the full dataset, the assigned JSD values range from 0.0093 to 0.3637, with a mean of 0.0506. As a practical guideline based on the validation results shown in Fig. \ref{F3}, spectra with assigned JSD $< 0.1$ (93.07\%) are highly reliable, those between 0.1 and 0.2 (6.24\%) remain acceptable for most purposes, while values exceeding 0.2 (0.69\%) indicate predictions that should be treated with greater caution. These per‑spectrum JSD values are included as ancillary metadata in the public data release.

\begin{figure}
\centering
\includegraphics[width=0.5\textwidth]{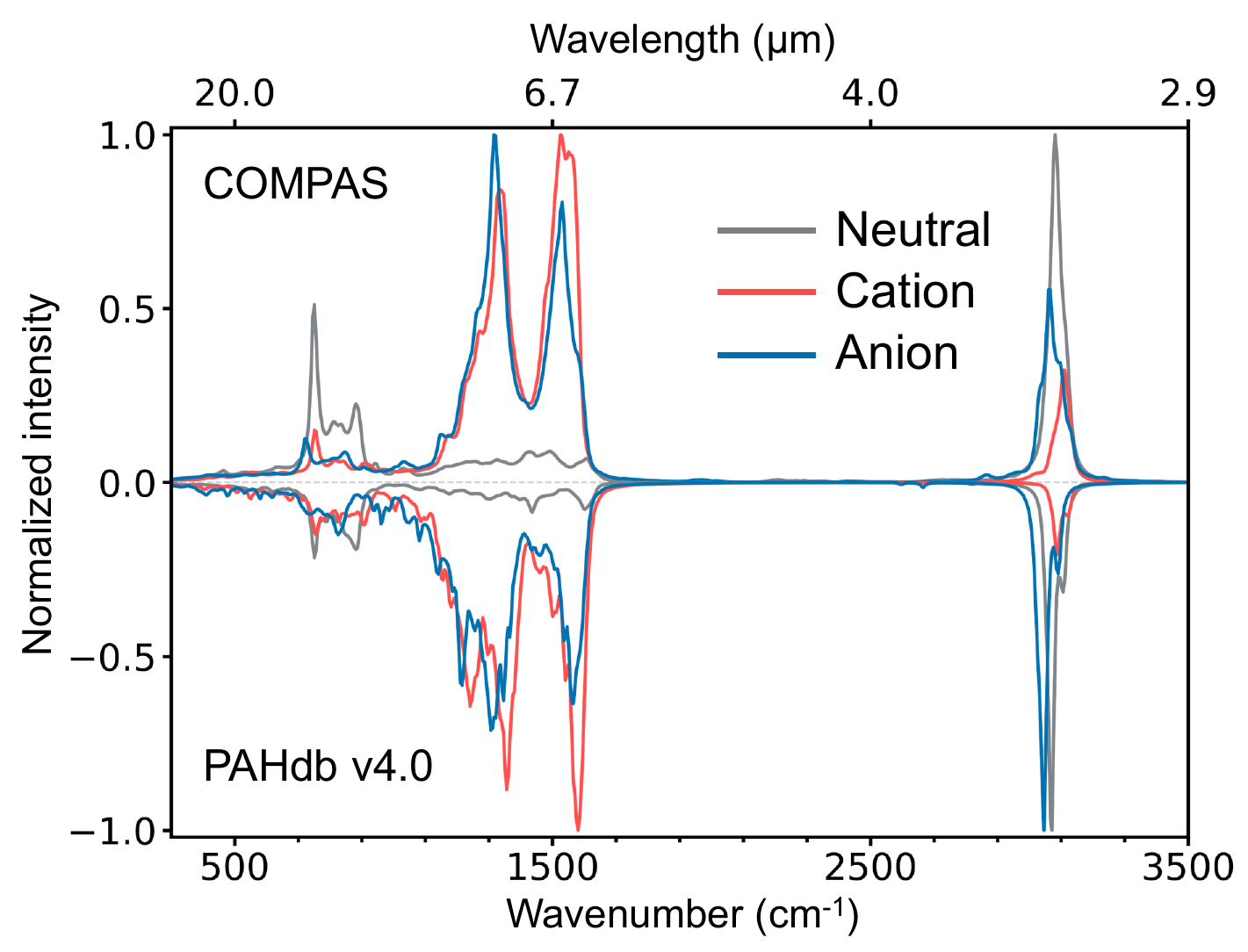}
\caption{Charge-state-dependent summed IR spectra of COMPAS PAHs predicted by ML-DH (above zero), compared with PAHdb reference spectra for molecules with $N_{\mathrm{C}} < 45$ that have spectral data available for all three charge states (below zero, mirrored). The spectra are normalized to unity separately for each charge state, so the comparison highlights the spectral shape rather than absolute intensities.}
\label{F6} 
\end{figure}

Figure~\ref{F6} presents the charge-state-dependent summed IR spectra of the 48,037 PAH structures from the COMPAS dataset predicted by ML-DH. For comparison, the corresponding summed spectra from PAHdb with molecular sizes of $N_{\mathrm{C}} < 45$ are shown as the mirrored spectra below the $y=0$ horizontal axis. Overall, the ML-DH summed spectra are in good agreement with the PAHdb reference spectra in terms of both band positions and relative intensities. The dominant vibrational features are consistently reproduced across all three charge states. The spectra also exhibit a strong charge-state dependence, which is characteristic of small to medium-sized PAHs: neutral species are dominated by the C--H stretching band near \SI{3.2}{\micro\meter}, while cationic and anionic PAHs show significantly enhanced mid-IR activity in the \SIrange{6}{9}{\micro\meter} region, accompanied by a reduction in the relative intensity of the C--H stretching band. This trend is consistent with the well-known charge dependence of PAH IR spectra.

\begin{figure}
\centering
\includegraphics[width=0.5\textwidth]{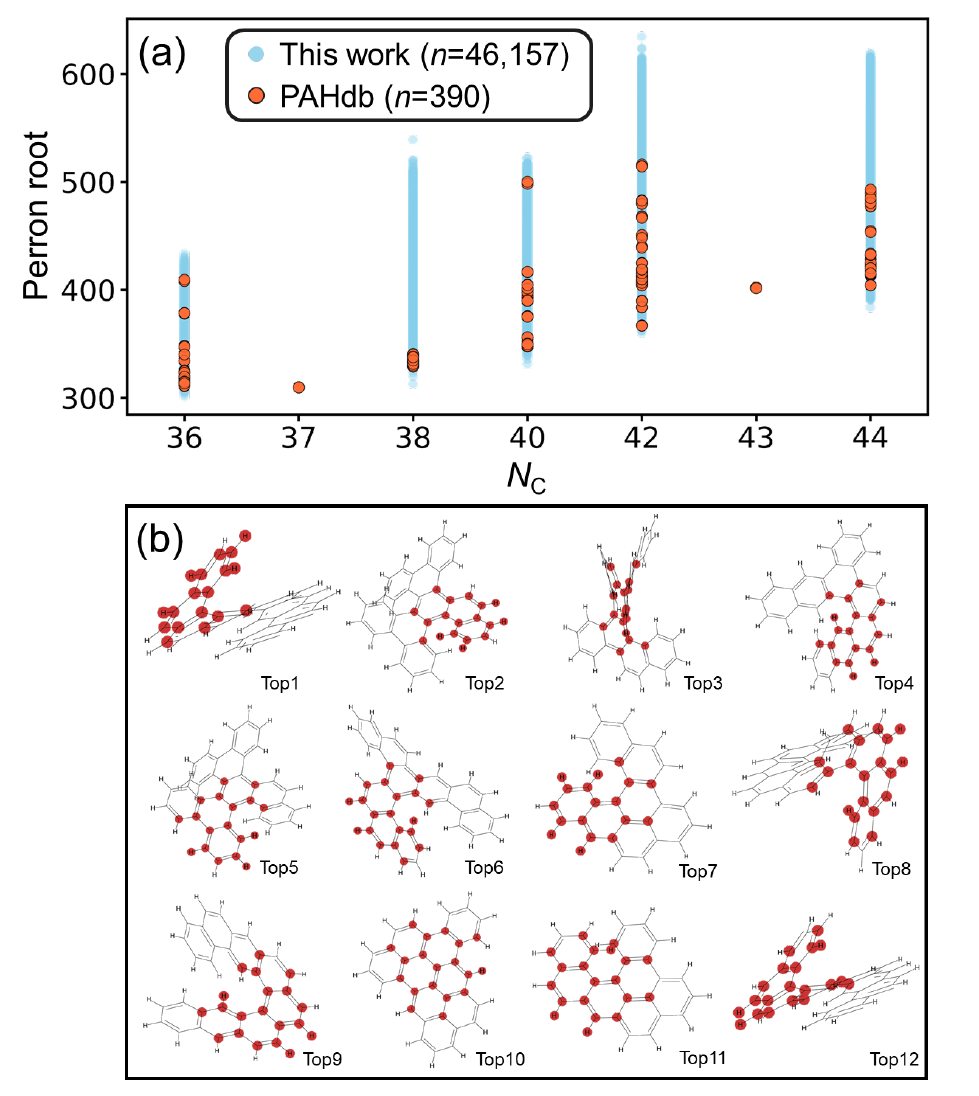}
\caption{Panel (a): Comparison of Perron root distributions of the COMPAS and PAHdb datasets as a function of $N_{\mathrm{C}}$ in the range $35$ to $45$. Panel (b): Top 12 molecular fingerprint fragments present in our dataset but absent from PAHdb. The red dots denote the fragment substructures, and the surrounding molecular frameworks are shown for context.}
\label{F7} 
\end{figure}

Although the summed spectra exhibit strong overall similarity to the PAHdb references, our dataset encompasses a diverse range of molecular structures that are absent from the existing database. To quantify this expanded coverage of the PAH chemical space, we computed the Perron root for all molecules in both datasets as a function of $N_{\mathrm{C}}$, as shown in Fig.~\ref{F7}a. The Perron root, defined as the maximum eigenvalue of the molecular distance matrix, is a commonly used descriptor of molecular structure, providing a first-order indication of structural diversity. For the size range $35 < N_{\mathrm{C}} < 45$, where the majority (96.1\%) of our PAHs reside, our dataset exhibits a dense and continuous distribution of Perron root values, whereas the PAHdb dataset shows only sparse and discrete sampling. This demonstrates that our dataset provides significantly more complete coverage of the PAH chemical space in this regime. 

To highlight the structural novelty of our dataset, we performed an extended‑connectivity fingerprint  analysis against PAHdb v4.0. By isolating molecular fragments unique to our collection, we identified a set of structural motifs absent from traditional databases. Figure~\ref{F7}b showcases 12 representative molecular fragments from this set, which predominantly feature nonplanar sterically crowded motifs or topologically complex fused-ring systems that are absent from PAHdb. These structures predominantly fall into two categories: (i) nonplanar, sterically crowded motifs, such as fjord, cove, and helical regions, where out-of-plane distortions arise from H--H stacking or steric congestion; and (ii) planar yet topologically complex fragments featuring curving-fused ring systems that are not represented in PAHdb. This structural expansion highlights the complementary nature of our dataset and its potential to probe previously unexplored regions of PAH chemical space in astrophysical contexts.

\subsection{Comparison with JWST/MIRI-MRS observations}

\begin{figure}
\centering
\includegraphics[width=0.5\textwidth]{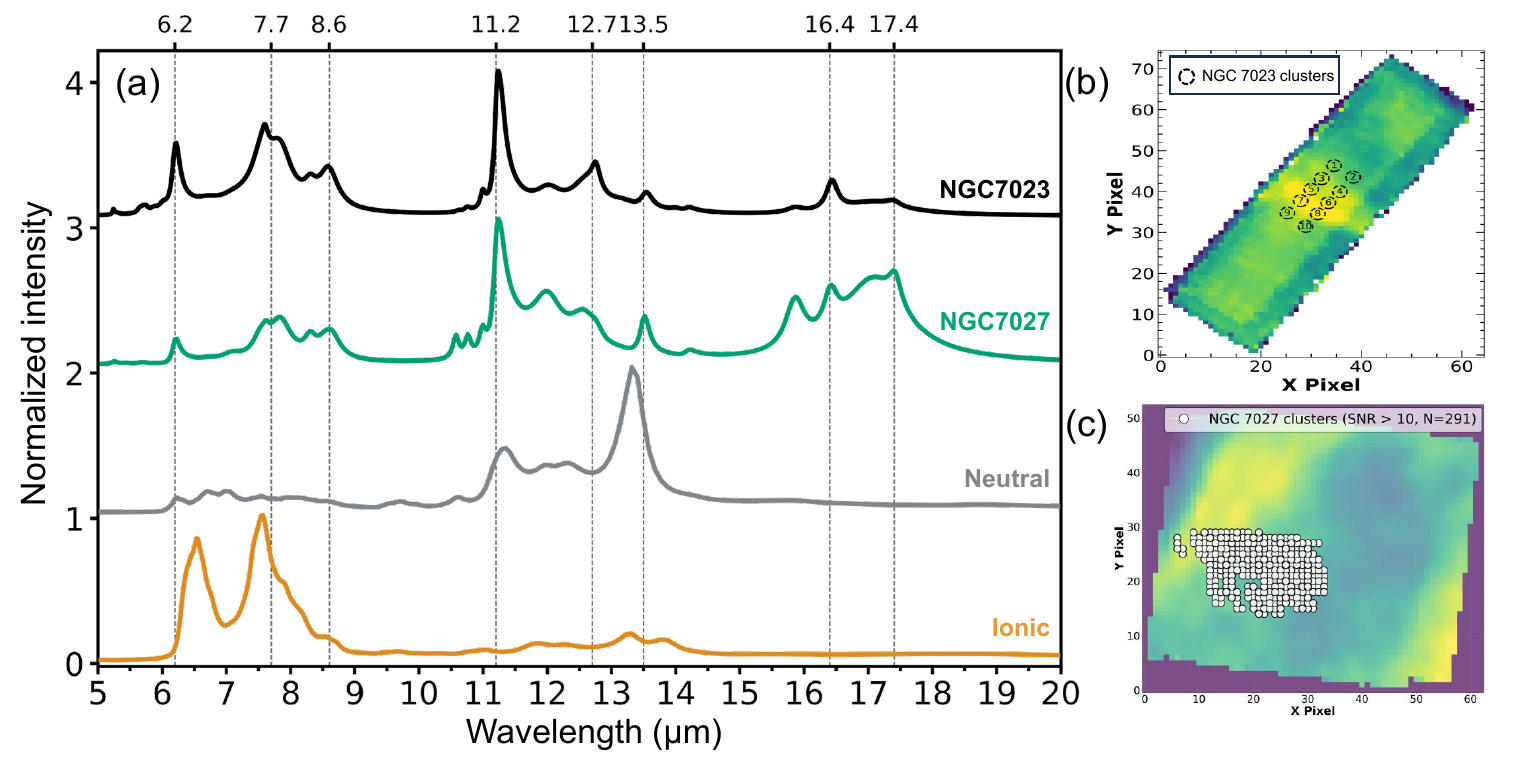}
\caption{Comparison between our spectral library and JWST/MIRI-MRS observations of NGC~7023 and NGC~7027. In panel (a), the two lower curves show synthetic total emission spectra computed using the cascade model implemented in the AmesPAHdbIDLSuite tool \citep{boersma2014} assuming a representative excitation energy of \SI{6}{\electronvolt} \citep{croiset2016}, based on the sum of all absorption spectra in our dataset, for neutral and ionic (cationic and anionic) species. The two upper curves show the NGC~7023 spectrum extracted from ten 0.5\arcsec-radius apertures (panel b) and the NGC~7027 spectrum averaged over PAH-emitting clumps with S/N~$>10$ in the 5--\SI{20}{\micro\meter} range (panel c). The total integrated flux is represented by the color intensity.}
\label{F8} 
\end{figure}

To demonstrate the astrophysical applicability of our synthetic library, we compared our composite spectra with observations of the Mid-Infrared Instrument-Medium Resolution Spectrometer (MIRI-MRS) on board JWST, focusing on the reflection nebula NGC~7023 and the planetary nebula NGC~7027 (Fig.~\ref{F8}). This comparison is intended as a qualitative sanity check and a demonstration of the potential of our spectral library, not as a quantitative fit to the observations or as a new physical explanation of the MIRI spectra. The composite spectra presented here are generated from a simple, equal-weight summation of all available spectra within each charge-state category, without any optimization of mixing ratios or detailed modeling of the local physical conditions (e.g., radiation field, PAH size distribution, or temperature). Because the resulting spectrum depends sensitively on the assumed excitation energy and mixing ratios, we do not interpret the relative contributions of different charge states or sizes in this section. We only compared the overall band positions and envelopes at a qualitative level. 

The NGC~7023 data (JWST Program~1192, PI: K. Misselt) and NGC~7027 data (JWST Program~1523, PI: D. R. Law) were reduced using the standard JWST Science Calibration Pipeline (v1.14.0), with additional custom corrections for instrumental artifacts, including pixel flagging and background subtraction; details are given in \citet{misselt2025}. The data cubes were resampled onto a common spatial grid with the \texttt{reproject} package \citep{Robitaille2020}, and spectra were extracted from 0.5\arcsec-radius circular apertures at ten positions across the photodissociation region (PDR) front of NGC~7023. NGC~7023 represents a prototypical PDR environment with well-stratified PAH size and charge distributions, while NGC~7027 exhibits a more processed PAH population. This comparison thus provides a valuable test of our library across different astrophysical conditions.

To generate synthetic emission spectra from our absorption spectra, we used the cascade model implemented in the AmesPAHdbIDLSuite tool \citep{boersma2014}. For this illustrative comparison, we assumed a representative single-photon excitation energy of 6 eV for all molecules, following the approach of \citet{croiset2016}. We note that this is a significant simplification. In reality, interstellar PAHs are exposed to a continuous distribution of radiation field energies, and the excitation energy strongly influences the partitioning of absorbed energy among different vibrational modes and, consequently, the relative intensities of different spectral bands. Higher excitation energies generally enhance the emission from high-frequency modes and can also affect the charge-state balance \citep{zhen2015,zhen2016,wenzel2020}. For NGC 7023 and NGC 7027, the radiation fields are distinctly different, and a single 6 eV excitation energy cannot capture this difference. A more realistic treatment would require modeling the full radiation field and the size-dependent absorption cross sections, which is beyond the scope of the present work.

The \SI{6.2}{\micro\meter} band is prominent in both sources. We note that our library does not include nitrogen-containing PAHs, which have been suggested as important carriers of this feature \citep{hudgins2005,boersma2013}, so no charge-state or carrier attribution is attempted here. The \SI{7.7}{\micro\meter} complex is the most intense PAH feature in both sources. Our composite spectra reproduce the overall position and width of this complex, but we do not use the relative contributions of different charge states in our equal-weight sum to draw conclusions about the ionization state of the observed PAHs. The 11.2 and \SI{12.7}{\micro\meter} bands are among the strongest features in both sources. Our composite spectra reproduce their overall positions, but we emphasize that the equal-weight sum does not allow us to infer the neutral versus ionic contribution in the observed sources.

Our composite spectra show no significant emission in the 15.6--\SI{18}{\micro\meter} region for either charge state under the equal-weight assumption. Both NGC~7027 and NGC~7023 exhibit clear features here, particularly the 16.4 and \SI{17.4}{\micro\meter} bands. This discrepancy may arise from: the largest PAHs ($N_{\mathrm{C}} > 50$) that are thought to contribute to this region \citep{boersma2010,ricca2012} are beyond our current size range; or, the emission may originate from more complex three-dimensional structures such as curved, fullerene-like skeletons \citep{Tan2026}. This highlights the complementary role of our dataset: while our library provides denser coverage in the small-to-medium size regime, it is best used alongside existing benchmarks such as PAHdb, which together offer a more complete view of the PAH spectral landscape.

We stress that the current comparison does not constrain the charge-state or size distribution of PAHs in NGC~7023 or NGC~7027. A rigorous interpretation would require detailed population synthesis or ML models that properly account for the local radiation field, ionization balance, and size-dependent excitation \citep{Wang2026}, which is beyond the scope of the present work. The value of our library at this stage lies in its dense coverage of the small-to-medium PAH chemical space, which complements existing benchmarks such as PAHdb.

\section{Conclusions}

We have presented a large-scale dataset of PAH IR spectra for astronomical applications, comprising 144,111 spectra across 48,037 PAH structures from the COMPAS dataset in neutral, cationic, and anionic charge states. By integrating neural network force fields and electronic dipole neural networks, our ML-DH framework offers a computationally efficient pathway to systematically expand the chemical space of available PAH reference spectra. The resulting dataset is not a replacement for existing libraries such as PAHdb, but rather serves as a companion resource that targets structural regimes, particularly nonplanar, sterically crowded, and topologically complex motifs, which have been sparsely sampled in previous databases. Rigorous validation against DFT benchmarks and PAHdb, along with per-molecule uncertainty estimates provided throughout the dataset, demonstrates the high reliability and transparency of our predictions. To support AIBs studies by JWST and other IR facilities, both the predicted IR spectra and the optimized molecular geometries, along with the model code, are made publicly available. Looking ahead, we plan to expand the chemical coverage to include heteroatom-bearing PAHs and open-shell species, while updating the database with spectra generated at higher levels of theory.

\section*{Data availability}

The full spectral dataset, including equilibrium geometries, calculated vibrational frequencies, IR intensities, and reliability metrics, is publicly available on Zenodo at \href{https://doi.org/10.5281/zenodo.21506061}{10.5281/zenodo.21506061}. The source code for generating this dataset is freely accessible on GitHub at Repository \href{https://github.com/zwAstroChem/ML-DH-PAH-IR}{ML-DH-PAH-IR}. It includes the already-trained models and can be used out of the box to rapidly generate IR spectra for PAHs with similar structures, with no further ML training required.

\bibliographystyle{aa}

\begin{thebibliography}{54}
\expandafter\ifx\csname natexlab\endcsname\relax\def\natexlab#1{#1}\fi

\bibitem[{Allamandola {et~al.}(1985)Allamandola, Tielens, \& Barker}]{allamandola1985}
Allamandola, L.~J., Tielens, A. G. G.~M., \& Barker, J.~R. 1985, ApJ, 290, L25

\bibitem[{Bauschlicher \& Langhoff(1997)}]{bauschlicher1997}
Bauschlicher, C.~W. \& Langhoff, S.~R. 1997, Spectrochim. Acta A, 53, 1225

\bibitem[{Bauschlicher~Jr. {et~al.}(2010)Bauschlicher~Jr., Boersma, Ricca, Mattioda, Cami, Peeters, S{\'a}nchez~de Armas, Puerta~Saborido, Hudgins, \& Allamandola}]{bauschlicher2010}
Bauschlicher~Jr., C.~W., Boersma, C., Ricca, A., {et~al.} 2010, ApJS, 189, 341

\bibitem[{Bauschlicher~Jr. {et~al.}(2018)Bauschlicher~Jr., Ricca, Boersma, \& Allamandola}]{bauschlicher2018}
Bauschlicher~Jr., C.~W., Ricca, A., Boersma, C., \& Allamandola, L.~J. 2018, ApJS, 234, 32

\bibitem[{Bhatia {et~al.}(2025)Bhatia, Rinke, \& Krej{\v{c}}{\'\i}}]{bhatia2025}
Bhatia, N., Rinke, P., \& Krej{\v{c}}{\'\i}, O. 2025, npj Comput. Mater., 11, 324

\bibitem[{Boersma {et~al.}(2010)Boersma, Bauschlicher~Jr., Allamandola, Ricca, Peeters, \& Tielens}]{boersma2010}
Boersma, C., Bauschlicher~Jr., C.~W., Allamandola, L., {et~al.} 2010, A\&A, 511, A32

\bibitem[{Boersma {et~al.}(2014)Boersma, Bauschlicher~Jr., Ricca, Mattioda, Cami, Peeters, de~Armas, Saborido, Hudgins, \& Allamandola}]{boersma2014}
Boersma, C., Bauschlicher~Jr., C.~W., Ricca, A., {et~al.} 2014, ApJS, 211, 8

\bibitem[{Boersma {et~al.}(2013)Boersma, Bregman, \& Allamandola}]{boersma2013}
Boersma, C., Bregman, J., \& Allamandola, L. 2013, ApJ, 769, 117

\bibitem[{Croiset {et~al.}(2016)Croiset, Candian, Bern{\'e}, \& Tielens}]{croiset2016}
Croiset, B., Candian, A., Bern{\'e}, O., \& Tielens, A. 2016, A\&A, 590, A26

\bibitem[{Dias(2004)}]{Dias2004}
Dias, J.~R. 2004, Aust. J. Chem., 57, 1039

\bibitem[{Dias(2007)}]{Dias2007}
Dias, J.~R. 2007, J. Chem. Inf. Model., 47, 707

\bibitem[{Frisch {et~al.}(2016)Frisch, Trucks, Schlegel, Scuseria, Robb, Cheeseman, Scalmani, Barone, Petersson, Nakatsuji, {et~al.}}]{Frisch2016}
Frisch, M.~J., Trucks, G.~W., Schlegel, H.~B., {et~al.} 2016, Gaussian 16, Gaussian, Inc., Wallingford, CT, USA

\bibitem[{Galliano {et~al.}(2008)Galliano, Madden, Tielens, Peeters, \& Jones}]{galliano2008}
Galliano, F., Madden, S.~C., Tielens, A.~G., Peeters, E., \& Jones, A.~P. 2008, ApJ, 679, 310

\bibitem[{Gastegger {et~al.}(2017)Gastegger, Behler, \& Marquetand}]{gastegger2017}
Gastegger, M., Behler, J., \& Marquetand, P. 2017, Chem. Sci., 8, 6924

\bibitem[{He {et~al.}(2026)He, Mai, \& Wang}]{he2026}
He, J., Mai, X., \& Wang, Z. 2026, A\&A, 708, A335

\bibitem[{Hudgins {et~al.}(2005)Hudgins, Bauschlicher~Jr, \& Allamandola}]{hudgins2005}
Hudgins, D.~M., Bauschlicher~Jr, C.~W., \& Allamandola, L. 2005, ApJ, 632, 316

\bibitem[{Kovacs {et~al.}(2020)Kovacs, Zhu, Carrete, Madsen, \& Wang}]{kovacs2020}
Kovacs, P., Zhu, X., Carrete, J., Madsen, G.~K., \& Wang, Z. 2020, ApJ, 902, 100

\bibitem[{Langhoff(1996)}]{langhoff1996}
Langhoff, S.~R. 1996, J. Phys. Chem., 100, 2819

\bibitem[{Larsen {et~al.}(2017)Larsen, Mortensen, Blomqvist, Castelli, Christensen, Du{\l}ak, Friis, Groves, Hammer, Hargus, {et~al.}}]{larsen2017}
Larsen, A.~H., Mortensen, J.~J., Blomqvist, J., {et~al.} 2017, J. Phys.: Condens. Matter, 29, 273002

\bibitem[{Leger \& Puget(1984)}]{leger1984}
Leger, A. \& Puget, J.~L. 1984, A\&A, 137, L5

\bibitem[{Li(2020)}]{li2020}
Li, A. 2020, Nat. Astron., 4, 339

\bibitem[{Li {et~al.}(2024)Li, Li, Yang, \& Fang}]{li2024}
Li, K., Li, A., Yang, X., \& Fang, T. 2024, MNRAS, 529, 4425

\bibitem[{Liu {et~al.}(2026)Liu, Wang, \& Qiu}]{liu2026}
Liu, Y., Wang, Z., \& Qiu, D. 2026, MNRAS, 549, stag893

\bibitem[{Mackie {et~al.}(2015)Mackie, Candian, Huang, Maltseva, Petrignani, Oomens, Buma, Lee, \& Tielens}]{mackie2015}
Mackie, C.~J., Candian, A., Huang, X., {et~al.} 2015, J. Chem. Phys., 143, 224314

\bibitem[{Mackie {et~al.}(2018)Mackie, Chen, Candian, Lee, \& Tielens}]{mackie2018}
Mackie, C.~J., Chen, T., Candian, A., Lee, T.~J., \& Tielens, A. G. G.~M. 2018, J. Chem. Phys., 149, 134302

\bibitem[{Mai {et~al.}(2025)Mai, Wang, Pan, Sch{\"o}rghuber, Kov{\'a}cs, Carrete, \& Madsen}]{mai2025}
Mai, X., Wang, Z., Pan, L., {et~al.} 2025, MNRAS, 541, 3073

\bibitem[{Maragkoudakis {et~al.}(2020)Maragkoudakis, Peeters, \& Ricca}]{maragkoudakis2020}
Maragkoudakis, A., Peeters, E., \& Ricca, A. 2020, MNRAS, 494, 642

\bibitem[{Mattioda {et~al.}(2020)Mattioda, Hudgins, Boersma, Bauschlicher~Jr., Ricca, Cami, Peeters, de~Armas, Saborido, \& Allamandola}]{mattioda2020}
Mattioda, A.~L., Hudgins, D.~M., Boersma, C., {et~al.} 2020, ApJS, 251, 22

\bibitem[{Mayo~Yanes {et~al.}(2024)Mayo~Yanes, Chakraborty, \& Gershoni-Poranne}]{mayo2024}
Mayo~Yanes, E., Chakraborty, S., \& Gershoni-Poranne, R. 2024, Sci. Data, 11, 97

\bibitem[{McGill {et~al.}(2021)McGill, Forsuelo, Guan, \& Green}]{mcgill2021}
McGill, C., Forsuelo, M., Guan, Y., \& Green, W.~H. 2021, J. Chem. Inf. Model., 61, 2594

\bibitem[{Meng {et~al.}(2023)Meng, Zhang, Liang, \& Wang}]{meng2023machine}
Meng, Z., Zhang, Y., Liang, E., \& Wang, Z. 2023, MNRAS, 525, L29

\bibitem[{Meng {et~al.}(2021)Meng, Zhu, Kov{\'a}cs, Liang, \& Wang}]{meng2021}
Meng, Z., Zhu, X., Kov{\'a}cs, P., Liang, E., \& Wang, Z. 2021, ApJ, 922, 101

\bibitem[{Metcalf {et~al.}(2021)Metcalf, Jiang, Spronk, Cheney, \& Sherrill}]{metcalf2021}
Metcalf, D.~P., Jiang, A., Spronk, S.~A., Cheney, D.~L., \& Sherrill, C.~D. 2021, J. Chem. Inf. Model., 61, 115

\bibitem[{Misselt {et~al.}(2025)Misselt, Witt, Gordon, Van De~Putte, Trahin, Abergel, Noriega-Crespo, Guillard, Zannese, Dellâ€™Ova, {et~al.}}]{misselt2025}
Misselt, K., Witt, A., Gordon, K., {et~al.} 2025, A\&A, 700, A158

\bibitem[{Montes-Campos {et~al.}(2021)Montes-Campos, Carrete, Bichelmaier, Varela, \& Madsen}]{montes-campos2021}
Montes-Campos, H., Carrete, J., Bichelmaier, S., Varela, L.~M., \& Madsen, G. K.~H. 2021, J. Chem. Inf. Model., 62, 88

\bibitem[{Neugebauer {et~al.}(2002)Neugebauer, Reiher, Kind, \& Hess}]{neugebauer2002}
Neugebauer, J., Reiher, M., Kind, C., \& Hess, B.~A. 2002, J. Comput. Chem., 23, 895

\bibitem[{Pan {et~al.}(2024)Pan, Carrete, Wang, \& Madsen}]{pan2024}
Pan, L., Carrete, J., Wang, Z., \& Madsen, G.~K. 2024, Phys. Rev. B, 109, 035417

\bibitem[{Peeters {et~al.}(2002)Peeters, Hony, Van~Kerckhoven, Tielens, Allamandola, Hudgins, \& Bauschlicher}]{peeters2002}
Peeters, E., Hony, S., Van~Kerckhoven, C., {et~al.} 2002, A\&A, 390, 1089

\bibitem[{Peeters {et~al.}(2021)Peeters, Mackie, Candian, \& Tielens}]{peeters2021}
Peeters, E., Mackie, C.~J., Candian, A., \& Tielens, A. G. G.~M. 2021, Acc. Chem. Res., 54, 1921

\bibitem[{Ricca {et~al.}(2012)Ricca, Bauschlicher~Jr., Boersma, Tielens, \& Allamandola}]{ricca2012}
Ricca, A., Bauschlicher~Jr., C.~W., Boersma, C., Tielens, A. G. G.~M., \& Allamandola, L.~J. 2012, ApJ, 754, 75

\bibitem[{Ricca {et~al.}(2026)Ricca, Boersma, Maragkoudakis, Roser, Shannon, Allamandola, \& Bauschlicher~Jr}]{ricca2026}
Ricca, A., Boersma, C., Maragkoudakis, A., {et~al.} 2026, ApJS, 282, 7

\bibitem[{Rigopoulou {et~al.}(2021)Rigopoulou, Barale, Clary, Shan, Alonso-Herrero, Garc{\'\i}a-Bernete, Hunt, Kerkeni, Pereira-Santaella, \& Roche}]{rigopoulou2021}
Rigopoulou, D., Barale, M., Clary, D., {et~al.} 2021, MNRAS, 504, 5287

\bibitem[{{Robitaille} {et~al.}(2020){Robitaille}, {Deil}, \& {Ginsburg}}]{Robitaille2020}
{Robitaille}, T., {Deil}, C., \& {Ginsburg}, A. 2020, {reproject: Python-based astronomical image reprojection}, Astrophysics Source Code Library, record ascl:2011.023

\bibitem[{Rogers \& Hahn(2010)}]{rogers2010}
Rogers, D. \& Hahn, M. 2010, J. Chem. Inf. Model., 50, 742

\bibitem[{Saquer {et~al.}(2024)Saquer, Iqbal, Ellis, \& Yoshimatsu}]{saquer2024}
Saquer, N., Iqbal, R., Ellis, J.~D., \& Yoshimatsu, K. 2024, Digit. Discov., 3, 602

\bibitem[{Tan {et~al.}(2026)Tan, Wang, Lu, Haidar, \& Rigopoulou}]{Tan2026}
Tan, X., Wang, Z., Lu, Z., Haidar, H., \& Rigopoulou, D. 2026, arXiv preprint [\eprint[arXiv]{2608.18505}]

\bibitem[{Tang {et~al.}(2026)Tang, He, Wang, \& Qiu}]{tang2026}
Tang, G., He, J., Wang, Z., \& Qiu, D. 2026, MNRAS, 546, stag283

\bibitem[{Tang {et~al.}(2023)Tang, Bromley, \& Hammer}]{tang2023}
Tang, Z., Bromley, S.~T., \& Hammer, B. 2023, J. Chem. Phys., 158, 224108

\bibitem[{Wahab \& Gershoni-Poranne(2024)}]{wahab2024}
Wahab, A. \& Gershoni-Poranne, R. 2024, Phys. Chem. Chem. Phys., 26, 15344

\bibitem[{Wahab {et~al.}(2022)Wahab, Pfuderer, Paenurk, \& Gershoni-Poranne}]{wahab2022}
Wahab, A., Pfuderer, L., Paenurk, E., \& Gershoni-Poranne, R. 2022, J. Chem. Inf. Model., 62, 3704

\bibitem[{{Wang}(2026)}]{Wang2026}
{Wang}, Z. 2026, \aap, 710, L17

\bibitem[{Wenzel {et~al.}(2020)Wenzel, Joblin, Giuliani, Rodriguez~Castillo, Mulas, Ji, Sabbah, Quiroga, Pe{\~n}a, \& Nahon}]{wenzel2020}
Wenzel, G., Joblin, C., Giuliani, A., {et~al.} 2020, A\&A, 641, A98

\bibitem[{Zhen {et~al.}(2015)Zhen, Castellanos, Paardekooper, Ligterink, Linnartz, Nahon, Joblin, \& Tielens}]{zhen2015}
Zhen, J., Castellanos, P., Paardekooper, D.~M., {et~al.} 2015, ApJL, 804, L7

\bibitem[{Zhen {et~al.}(2016)Zhen, Castillo, Joblin, Mulas, Sabbah, Giuliani, Nahon, Martin, Champeaux, \& Mayer}]{zhen2016}
Zhen, J., Castillo, S.~R., Joblin, C., {et~al.} 2016, ApJ, 822, 113

\end{thebibliography}

\FloatBarrier 
\clearpage
\end{document}